\documentclass[
reprint,
superscriptaddress,
amsmath,
amssymb,
aps,
longbibliography,
pra,
showpacs,
floatfix
]{revtex4-1}

\usepackage{graphicx}
\usepackage{color}

\usepackage{tabularx}
\usepackage{multirow}
\usepackage{dcolumn} 
\newcolumntype{d}[1]{D{.}{.}{#1}}
\newcolumntype{L}[1]{>{\raggedright\arraybackslash}p{#1}}
\newcolumntype{C}[1]{>{\centering\arraybackslash}p{#1}}
\newcolumntype{R}[1]{>{\raggedleft\arraybackslash}p{#1}}
\usepackage{booktabs}

\usepackage{comment}
\usepackage{enumitem}

\usepackage{hyperref}
\definecolor{color1}{rgb}{0,0.25,0.70}
\hypersetup{colorlinks=true, 
    linkcolor={color1},
    citecolor={color1}, 
    urlcolor={color1}
}

\newcommand{\ang}{\ensuremath{\mathrm{\AA}}}

\begin{document}

\title{Phono-magnetic analogs to opto-magnetic effects}

\author{Dominik~M.\ Juraschek}
\email{djuraschek@seas.harvard.edu}
\affiliation{Harvard John A. Paulson School of Engineering and Applied Sciences, Harvard University, Cambridge, MA 02138, USA}
\affiliation{Department of Materials, ETH Zurich, CH-8093 Z\"{u}rich, Switzerland}
\author{Prineha Narang}
\affiliation{Harvard John A. Paulson School of Engineering and Applied Sciences, Harvard University, Cambridge, MA 02138, USA}
\author{Nicola~A.\ Spaldin}
\affiliation{Department of Materials, ETH Zurich, CH-8093 Z\"{u}rich, Switzerland}

\date{\today}

\begin{abstract}
The magneto-optical and opto-magnetic effects describe the interaction of light with a magnetic medium. The most prominent examples are the Faraday and Cotton-Mouton effects that modify the transmission of light through a medium, and the inverse Faraday and inverse Cotton-Mouton effects that produce effective magnetic fields for the spin in the material. Here, we introduce the phenomenology of the analogous magneto-\textit{phononic} and \textit{phono}-magnetic effects, in which vibrational quanta take the place of the light quanta. We show, using a combination of first-principles calculations and phenomenological modeling, that the effective magnetic fields exerted by the phonon analogs of the inverse Faraday and inverse Cotton-Mouton effects on the spins of antiferromagnetic nickel oxide yield magnitudes comparable to and potentially larger than those of the opto-magnetic originals.
\end{abstract}

\maketitle



Antiferromagnets are a promising alternative to ferromagnets in spintronics applications, such as magnetic recording and data storage, because of their orders-of-magnitude faster terahertz operations compared to gigahertz speeds for ferromagnets \cite{Jungwirth2016,Baltz2018,Jungwirth2018}. To achieve these operation speeds, ultrashort light pulses are used to manipulate the magnetic order on timescales of less than a picosecond \cite{Beaurepaire1996,Kirilyuk2010,Kalashnikova2015,Nemec2018}. One ingredient of ultrafast spin control is the excitation of coherent spin waves (magnons) that can be achieved through opto-magnetic effects, in which light couples to the spins directly \cite{Kampfrath2011,Baierl2016_2}, or through modulations of the dielectric function \cite{kimel:2005,Kalashnikova2007,Kalashnikova2008,Gridnev2008,Satoh2017,Tzschaschel2017}, exchange interactions \cite{Kubacka2014}, and anisotropy \cite{Baierl2016,Schlauderer2019}. At the same time, the coupling of vibrations of the crystal lattice (phonons) to the spins has long been known to influence the magnetic order by compensating the angular momentum in spin relaxation \cite{VanVleck1940,Garanin2015,Nakane2018,Lunghi2019} and demagnetization dynamics \cite{zhang:2014,Henighan2016_2,Dornes2019,Mentink2019}. Recent experiments have shown that, when coherently excited with a terahertz pulse, infrared-active phonon modes transfer energy to the magnetic order through the spin-phonon interaction. This leads to coherent spin-wave excitations \cite{nova:2017} and is capable of inducing and reorienting magnetic order \cite{Disa2020,Afanasiev2019}. Theoretical explanations of these phenomena have so far taken into account dynamical magnetic fields \cite{juraschek2:2017,Juraschek2019} and transient distortions of the crystal lattice \cite{juraschek:2017,Radaelli2018,Fechner2018,Gu2018,Khalsa2018,Maehrlein2018,Rodriguez-Vega2020} produced by the coherent phonon excitation. These developments mark the emergence of a new field of phono-magnetism that is complementary to the established opto-magnetism approach to antiferromagnetic spintronics and requires an equivalent theoretical foundation.

\begin{figure}[t]
\centering
\includegraphics[scale=0.15]{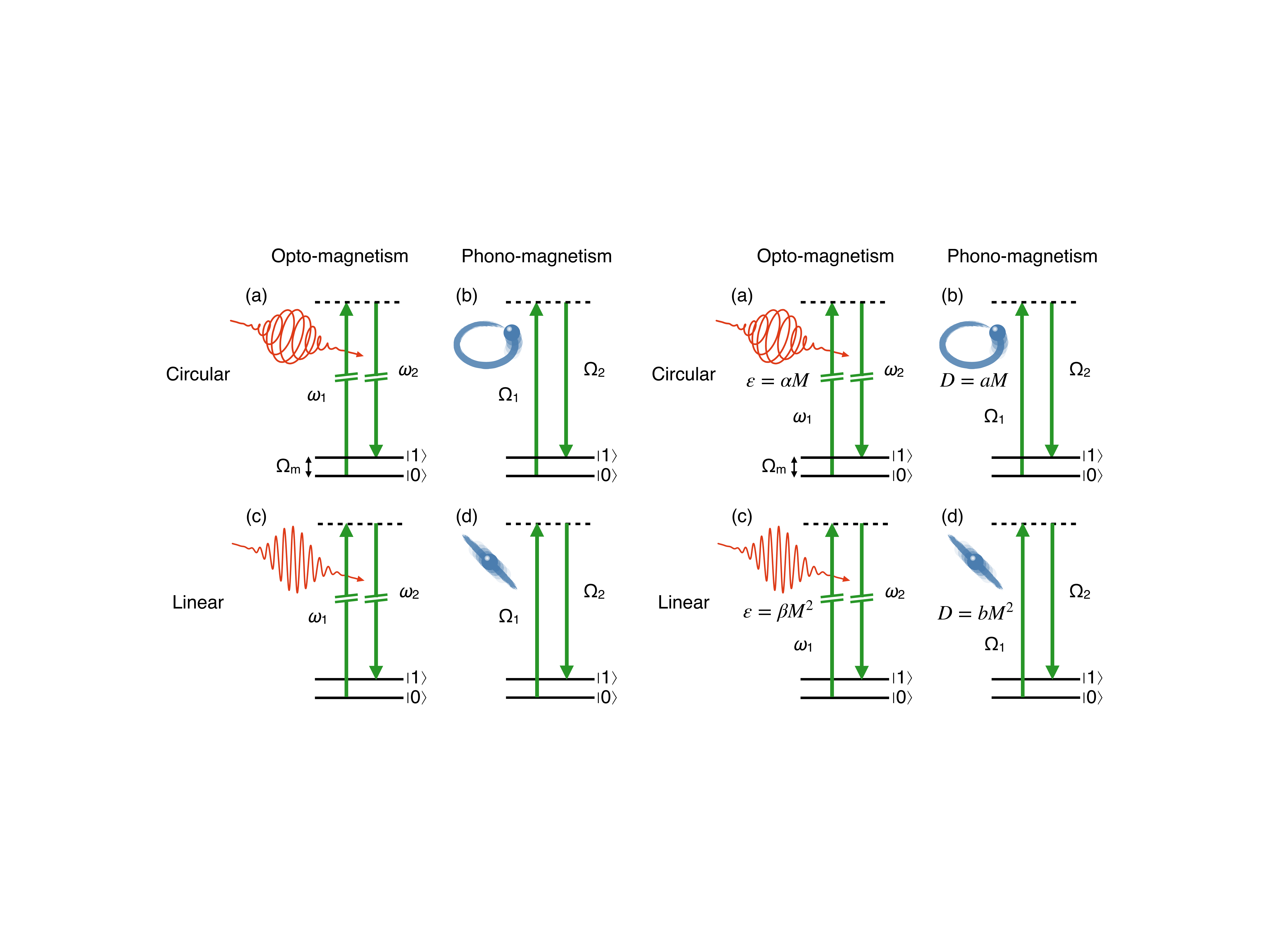}
\caption{
Opto-magnetic and phono-magnetic effects mediated by circularly and linearly polarized photons and phonons. The difference between two frequency components of an ultrashort laser pulse, $\omega_1$ and $\omega_2$, or two coherently excited phonons, $\Omega_1$ and $\Omega_2$, are resonant with one or two magnons with frequency $\Omega_m$. The fundamental process for photons is impulsive stimulated Raman scattering; for phonons it is ionic Raman scattering. (a) Inverse Faraday effect, (b) phonon inverse Faraday effect, (c) inverse Cotton-Mouton effect, (d) phonon inverse Cotton-Mouton effect.
}
\label{fig:optophonomagnetism}
\end{figure}

The purpose of this study is to develop a theoretical framework for phono-magnetic phenomena that is based on the analogy to opto-magnetic effects. We begin by showing that the magneto-optical Faraday and Cotton-Mouton and the opto-magnetic inverse effects have magneto-phononic and phono-magnetic analogs when we replace photons by phonons in the interaction, as illustrated in Fig.~\ref{fig:optophonomagnetism}. We next calculate the corresponding interaction strengths for the paradigmatic antiferromagnet nickel oxide (NiO) from first principles and estimate the effective magnetic fields produced by the opto-magnetic inverse Faraday and inverse Cotton-Mouton effects and their phono-magnetic analogs as a result of coherent optical and phononic driving. The opto-magnetic effects that we consider here are fundamentally impulsive stimulated Raman scattering processes, in which the energy of light is transferred to the magnetic system through intermediate virtual or real electronic states. The phono-magnetic effects that we consider here are ionic Raman scattering processes, in which the energy of light is transferred to coherently excited optical phonons, which in turn transfer their energy to the magnetic system through spin-phonon coupling. Ionic Raman scattering was predicted half a century ago \cite{maradudin:1970,Wallis1971,Humphreys1972} and has been demonstrated recently for nonlinear phonon interactions \cite{forst:2011,subedi:2014,Mankowsky:2015}. The approach presented here can be straightforwardly extended for non-Raman mechanisms, such as one-photon and one-phonon \cite{Fransson2017,Hashimoto2017,Roychoudhury2018,Fechner2018,Nomura2019,Streib2018,Hellsvik2019,Maldonado2019,Berk2019,Streib2019,Ruckriegel2020}, as well as two-magnon processes \cite{Bossini2016,Bossini2019}, which will be the matter of future work. Our results pave the way for future experiments utilizing phono-magnetic effects in the ultrafast control of magnetic order.


\section{Theoretical formalism}

We treat the coupling of coherent light or phonons with the spins in the effective magnetic field picture that provides an intuitive description of the physics and has commonly been applied to the opto-magnetic effects in the literature \cite{Pershan1967,Kirilyuk2010,Kalashnikova2015}. The description is valid in the regime of excitations in which no photo- or phonon-induced melting of magnetic order occurs \cite{Rini2007,Tobey2008,Beaud_et_al:2009,Caviglia2013,Hu2016,Forst2017}. We will comment on the limitations of this formalism in the discussion section.



\subsection{Opto-magnetism}

We begin by reviewing the classical interaction of light with magnetic matter, whose Hamiltonian, $\mathcal{H}^\mathrm{phot}$, can be written as \cite{Kalashnikova2015,Tzschaschel2017}
\begin{equation}\label{eq:light-matter}
\mathcal{H}^\mathrm{phot}=\tilde{\varepsilon}_{ij}(\mathbf{M},\mathbf{L})E_iE_j^\ast,
\end{equation}
where $\tilde{\varepsilon}_{ij}=-\varepsilon_{ij}V_\mathrm{c}$, $\varepsilon_{ij}$ is the frequency-dependent dielectric tensor, $V_\mathrm{c}$ is the volume of the unit cell, $E_i$ and $E_j^\ast$ are the complex electric-field components of light, and the indices $i$ and $j$ denote spatial coordinates. (We use the Einstein notation for summing indices.) In a general antiferromagnet with two magnetic sublattices, the dielectric tensor is a complex function of the ferro and antiferromagnetic vectors $\mathbf{M}=\mathbf{M}_1+\mathbf{M}_2$ and $\mathbf{L}=\mathbf{M}_1-\mathbf{M}_2$ that consist of the sum and the difference of the sublattice magnetizations $\mathbf{M}_1$ and $\mathbf{M}_2$, respectively. An expansion of $\tilde{\varepsilon}_{ij}$ up to second order in $\mathbf{M}$ and $\mathbf{L}$ yields \cite{Gridnev2008,Kalashnikova2015,Tzschaschel2017}:
\begin{eqnarray}
\tilde{\varepsilon}_{ij} & = & \tilde{\varepsilon}_{ij}^{(0)} + i\tilde{\alpha}_{ijk} M_k + i\tilde{\alpha}_{ijk}' L_k \nonumber \\
& + & \tilde{\beta}_{ijkl} M_k M_l +  + \tilde{\beta}_{ijkl}' L_k L_l + \tilde{\beta}_{ijkl}'' M_k L_l,
\end{eqnarray}
where $\tilde{\alpha}^{(}{'}{}^{)}$ and $\tilde{\beta}^{(}{'}{}^{,}{''}{}^{)}$ are the first and second-order magneto-optical coefficients with $\tilde{\alpha}^{(}{'}{}^{)}=\alpha^{(}{'}{}^{)}V_c$ and $\tilde{\beta}^{(}{'}{}^{,}{''}{}^{)}=\beta^{(}{'}{}^{,}{''}{}^{)}V_c$. The different magneto-optical and opto-magnetic effects can be classified according to their role in the expansion of the dielectric tensor, and we show three examples that are relevant for our study in Table~\ref{tab:light-matter}, adapted from Ref.~\cite{Kalashnikova2015}.

Magneto-optical effects occur as a result of the modulation of the dielectric function by magnetic order or fluctuations in the material and lead to circular birefringence, as described by the Faraday effect (first order in the magnetizations), and linear birefringence, as described by the Cotton-Mouton effect (second order in the magnetizations).

In the inverse opto-magnetic effects, the electric-field component of light acts as an effective magnetic field for the spins in the material, which is given by $\mathbf{B}^\mathrm{eff} = -V_\mathrm{c}^{-1}\partial \mathcal{H}^\mathrm{phot} / (\partial \mathbf{M})$ or $-V_\mathrm{c}^{-1} \partial \mathcal{H}^\mathrm{phot} / (\partial \mathbf{L})$. We look at the case of a coherent electric-field component that is provided by an ultrashort laser pulse, for which the effective magnetic field coherently excites magnons via impulsive stimulated Raman scattering involving two photons and one magnon \cite{Cottam1975,kimel:2005,Kalashnikova2007,Kalashnikova2008,reid:2010,popova:2011,Popova2012,Kalashnikova2015,Satoh2017,Tzschaschel2017}. To first order in the magnetizations, $\alpha_{ijk}=-\alpha_{jik}$, and we obtain the well-known form for the $z$-component of the effective magnetic field in the inverse Faraday effect, $B^\mathrm{eff}_z=i\alpha_{xyz}(E_x E_y^\ast - E_y E_x^\ast)=i\alpha_{xyz}(\mathbf{E}\times\mathbf{E}^\ast)_z$ \cite{vanderziel:1965,Pershan1967}. $B^\mathrm{eff}_z M_z$ then describes the coupling of spin angular momentum of circularly polarized light to the magnetization of the material, see Fig.~\ref{fig:optophonomagnetism}(a). In the inverse Cotton-Mouton effect (second order in the magnetizations), the light can be linearly polarized, as in Fig.~\ref{fig:optophonomagnetism}(c).


\begin{table*}[t]
\centering
\bgroup
\def\arraystretch{2.0}
\caption{
Classification of magneto-optical and opto-magnetic effects, adapted from Ref.~\cite{Kalashnikova2015}. Shown are the contributions to the interaction Hamiltonian $\mathcal{H}^\mathrm{phot}$, the components of the expansion of the volume dielectric tensor $\tilde{\varepsilon}_{ij}=-\varepsilon_{ij}V_\mathrm{c}$, and the effective magnetic fields $B^\mathrm{eff}$. Indices $i$, $j$, $k$, and $l$ denote spatial coordinates.
}
\begin{tabularx}{1.0\textwidth}{|C{3.93cm}|C{3.3cm}|C{3.3cm}|C{3.3cm}|C{3.3cm}|}
\hline
\parbox{3cm}{Contribution to \\ Hamiltonian} & \multicolumn{2}{c}{Magneto-optical effect} & \multicolumn{2}{|c|}{Opto-magnetic effect}\\[0.2cm]
\hline
$i\tilde{\alpha}_{ijk} E_i E_j^\ast M_k$ & $\tilde{\varepsilon}_{ij} = i\tilde{\alpha}_{ijk} M_k$ & Faraday effect & $B^\mathrm{eff}_k = i\alpha_{ijk} E_i E_j^\ast$ & \parbox{3cm}{Inverse \\ Faraday effect} \\[0.3cm]
\hline
$\tilde{\beta}_{ijkl} E_i E_j^\ast M_k M_l $ & $\tilde{\varepsilon}_{ij} = \tilde{\beta}_{ijkl} M_k M_l$ & \multirow{2}{*}{\parbox{3cm}{\vspace{0.5cm} Cotton-Mouton \\ effect}} & $B^\mathrm{eff}_l = \beta_{ijkl} E_i E_j^\ast M_k$ & \multirow{2}{*}{\parbox{3.3cm}{\vspace{0.5cm} Inverse \\ Cotton-Mouton \\ effect}} \\[0.3cm]
\cline{1-2}\cline{4-4}
$\tilde{\beta}'_{ijkl} E_i E_j^\ast L_k L_l $ & $\tilde{\varepsilon}_{ij} = \tilde{\beta}'_{ijkl} L_k L_l$ & & $B^\mathrm{eff}_l = \beta'_{ijkl} E_i E_j^\ast L_k$ & \\[0.3cm]
\hline
\end{tabularx}
\label{tab:light-matter}
\egroup
\end{table*}


\subsection{Phono-magnetism}

We now introduce an analogous description for the interaction of spins with lattice vibrations, in which \textit{phonons} take the place of photons. The normal coordinate (or amplitude) $Q$ of the phonon mode then takes the place of the electric-field component of the photon, and the interaction Hamiltonian, $\mathcal{H}^\mathrm{phon}$, can be written as
\begin{equation}\label{eq:phonon-matter}
\mathcal{H}^\mathrm{phon}=D_{ij}(\mathbf{M},\mathbf{L})Q_i{Q}^\ast_j.
\end{equation}
Here, the projected dynamical matrix $D_{ij}$ takes the place of the dielectric tensor as the coefficient. It is given by $D_{ij}= \mathbf{q}_i^\mathrm{T} D \mathbf{q}_j$, where $D$ is the dynamical matrix and $\mathbf{q}_{i/j}$ are the phonon eigenvectors, with indices $i$ and $j$ denoting the band number of the phonon mode. The case $i=j$ returns the eigenfrequency of phonon mode $i$, $D_{ii} = \Omega_i^2$. $Q$ is given in units of $\ang\sqrt{\mathrm{amu}}$, where amu is the atomic mass unit. The projected dynamical matrix can, analogously to the dielectric tensor, be expanded as a complex function of $\mathbf{M}$ and $\mathbf{L}$:
\begin{eqnarray}
D_{ij} & = & D_{ij}^{(0)} + i\tilde{a}_{ijk} M_k + i\tilde{a}_{ijk}' L_k \nonumber \\
& + & \tilde{b}_{ijkl} M_k M_l + \tilde{b}_{ijkl}' L_k L_l + \tilde{b}_{ijkl}'' M_k L_l,
\end{eqnarray}
where $\tilde{a}^{(}{'}{}^{)}$ and $\tilde{b}^{(}{'}{}^{,}{''}{}^{)}$ are the first and second-order magneto-phononic coefficients with $\tilde{a}^{(}{'}{}^{)}=a^{(}{'}{}^{)}V_c$ and $\tilde{b}^{(}{'}{}^{,}{''}{}^{)}=b^{(}{'}{}^{,}{''}{}^{)}V_c$. We classify the different magneto-phononic and phono-magnetic effects according to their role in the expansion of the projected dynamical matrix, and we show three examples that are relevant for our study in Table~\ref{tab:phonon-matter}.

Magneto-phononic effects occur as a result of the modulation of the interatomic forces, and therefore the dynamical matrix, by the magnetic order and magnetic fluctuations in the material through various types of spin-phonon interactions. Different manifestations of the magneto-phononic effects have been described in the literature throughout the past century: The most well-known first-order magneto-phononic effect is the Raman process in spin relaxation, in which a spin relaxes under the absorption and emission of two incoherent (thermal) phonons \cite{VanVleck1940,Orbach1961}. More recently identified first-order effects are the splitting of phonon modes in paramagnets in an external magnetic field \cite{schaack:1975,liu:2017}, the phonon Hall \cite{strohm:2005,sheng:2006,Zhang2011} and phonon Einstein-de Haas \cite{zhang:2014} effects in paramagnetic dielectrics, as well as for the phonon Zeeman effect in nonmagnetic dielectrics \cite{juraschek2:2017,Juraschek2019}. A manifestation of the second-order effect in the phenomenology can be found in the ionic contribution to the known magneto\textit{di}electric effect \cite{Lawes2009,Dubrovin2019}, in which the ionic part of the dielectric function, and therefore the dynamical matrix, is modulated by the magnetic order. Because of the analogy to the magneto-optical effects, these effects have been discussed as the Faraday and Cotton-Mouton effects of phonons in early studies \cite{anastassakis:1972,thalmeier:1978}.

We now propose that in the inverse phono-magnetic effects the phonon modes act as an effective magnetic field on the spins in the material, which is given by $\mathbf{B}^\mathrm{eff} = -V_\mathrm{c}^{-1}\partial \mathcal{H}^\mathrm{phon} / (\partial \mathbf{M})$ or $-V_\mathrm{c}^{-1}\partial \mathcal{H}^\mathrm{phon} / (\partial \mathbf{L})$. We look at the case of phonon modes that are coherently excited by an ultrashort pulse in the terahertz or mid-infrared spectral range, which in turn coherently excite magnons via ionic Raman scattering. The mechanism to first order of the magnetizations is the phonon analog to the inverse Faraday effect, and we therefore call it \textit{phonon inverse Faraday effect}. Here, $a_{ijk}=-a_{jik}$, and we obtain for the $z$-component of the effective magnetic field for the two components, $i$ and $j$, of a circularly polarized phonon in the $xy$-plane, $B^\mathrm{eff}_z=ia_{ijz}(Q_i Q_j^\ast - Q_j Q_i^\ast)=ia_{ijz}(\mathbf{Q}\times\mathbf{Q}^\ast)_z$. $B^\mathrm{eff}_z M_z$ then describes the coupling of phonon angular momentum with the magnetization of the material, see Fig.~\ref{fig:optophonomagnetism}(b). In the time domain, we obtain the more common form of this type of spin-phonon coupling, $B^\mathrm{eff}_z M_z=\bar{a}_{ijz}(\mathbf{Q}\times\dot{\mathbf{Q}})_z M_z$, where $\bar{a}_{ijz}=a_{ijz}/\Omega$ \cite{Ray1967,Rebane:1983,Ioselevich1995,sheng:2006,zhang:2014,juraschek2:2017,Juraschek2019}. The mechanism in second order in the magnetizations is the phonon analog to the inverse Cotton-Mouton effect, which we analogously call \textit{phonon inverse Cotton-Mouton effect}, and in which the phonons can be linearly polarized as in Fig.~\ref{fig:optophonomagnetism}(d).


\begin{table*}[t]
\centering
\bgroup
\def\arraystretch{2.0}
\caption{
Classification of magneto-phononic and phono-magnetic effects. Shown are the contributions to the interaction Hamiltonian $\mathcal{H}^\mathrm{phon}$, the components of the expansion of the projected dynamical matrix $D_{ij}$, and the effective magnetic fields $B^\mathrm{eff}$. Indices $i$ and $j$ denote the band index of the phonon modes, indices $k$, $l$ denote spatial coordinates.
}
\begin{tabularx}{1.0\textwidth}{|C{3.93cm}|C{3.3cm}|C{3.3cm}|C{3.3cm}|C{3.3cm}|}
\hline
\parbox{3cm}{Contribution to \\ Hamiltonian} & \multicolumn{2}{c}{Magneto-phononic effect} & \multicolumn{2}{|c|}{Phono-magnetic effect}\\[0.2cm]
\hline
$i\tilde{a}_{ijk} Q_i Q^\ast_j M_k$ & $D_{ij} = i\tilde{a}_{ijk} M_k$ & \parbox{3cm}{Phonon Faraday effect} & $B^\mathrm{eff}_k = ia_{ijk} Q_i Q^\ast_j$ & \parbox{3cm}{Phonon inverse \\ Faraday effect} \\[0.3cm]
\hline
$\tilde{b}_{ijkl} Q_i Q_j^\ast M_k M_l$ & $D_{ij} = \tilde{b}_{ijkl} M_k M_l$ & \multirow{2}{*}{\parbox{3cm}{\vspace{0.5cm} Phonon Cotton-Mouton\\ effect}} & $B^\mathrm{eff}_l = b_{ijkl} Q_i Q_j^\ast M_k$ & \multirow{2}{*}{\parbox{3.3cm}{\vspace{0.5cm} Phonon inverse \\ Cotton-Mouton \\ effect}} \\[0.3cm]
\cline{1-2}\cline{4-4}
$\tilde{b}_{ijkl}' Q_i Q_j^\ast L_k L_l $ & $D_{ij} = \tilde{b}_{ijkl}' L_k L_l$ & & $B^\mathrm{eff}_l = b_{ijkl}' Q_i Q_j^\ast L_k$ & \\[0.3cm]
\hline
\end{tabularx}
\label{tab:phonon-matter}
\egroup
\end{table*}


\begin{figure}[b]
\centering
\includegraphics[scale=0.068]{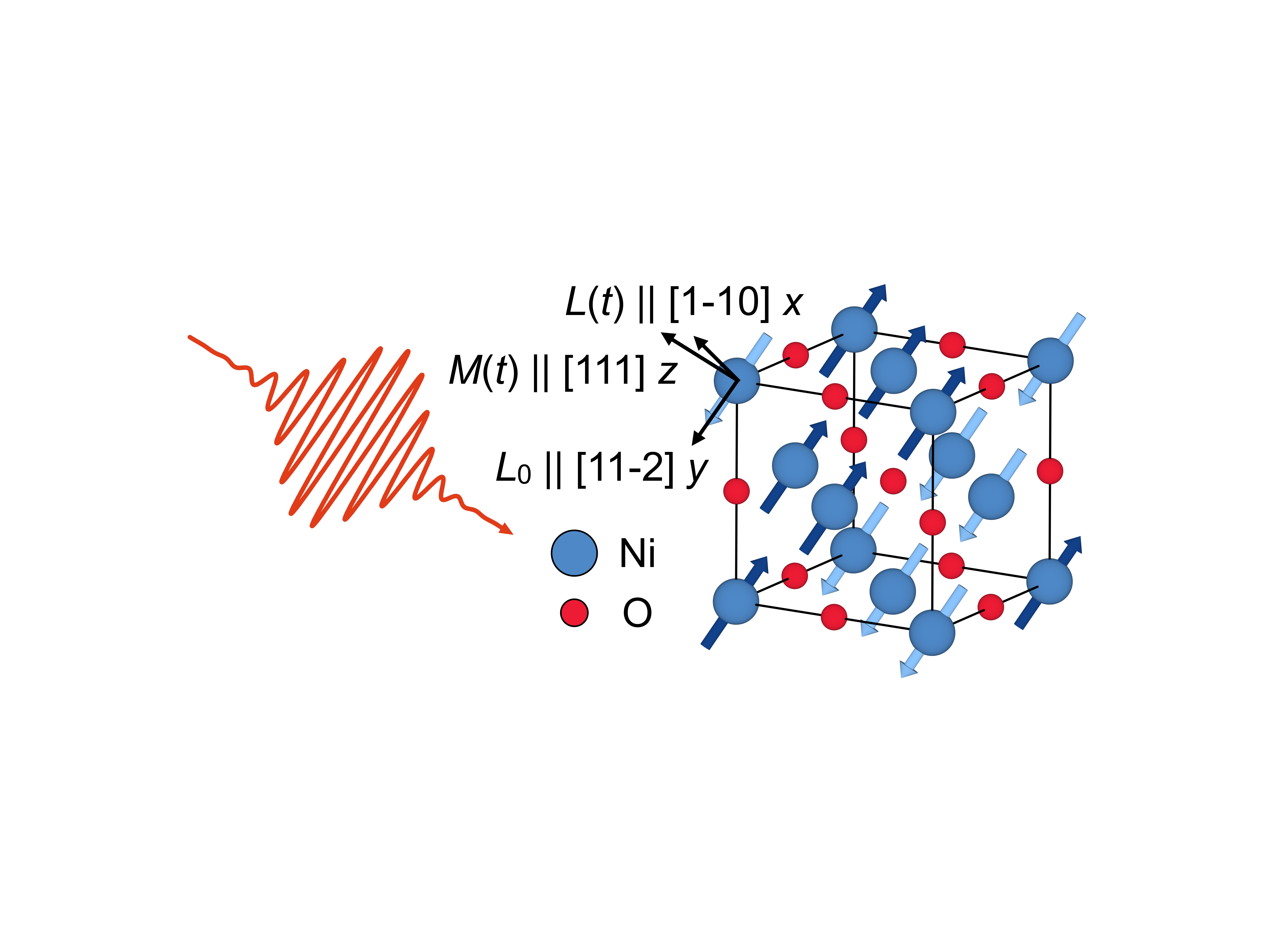}
\caption{
Antiferromagnetic order of NiO. The spins are aligned along one of three equivalent [11-2] directions. The in-plane magnon mode at 0.14~THz changes the antiferromagnetic vector, $L(t)$, along the [1-10] and the ferromagnetic vector, $M(t)$, along the [111] direction. $L_0$ is the equilibrium antiferromagnetic vector along the [11-2] direction.
}
\label{fig:niogeometry}
\end{figure}


\section{Results for Nickel oxide}

We turn to evaluating and estimating the strength of the opto-magnetic and proposed phono-magnetic effects for the example of NiO, in which coherent magnons have recently been generated via the inverse Faraday and inverse Cotton-Mouton effects \cite{Satoh2017,Tzschaschel2017}. 

In the paramagnetic phase, NiO crystallizes in the rocksalt structure (point group m$\bar{3}$m). Below the N\'{e}el temperature of $T_\mathrm{N}= 523~\mathrm{K}$, the spins couple ferromagnetically in the \{111\} planes of the crystal, with neighbouring planes of opposite ferromagnetic orientation, yielding a rhombohedrally distorted antiferromagnet with $\bar{3}$m point group \cite{Roth1960}. Within the \{111\} planes, the spins are aligned along one of three equivalent $\langle$11-2$\rangle$ directions, resulting in a further small monoclinic distortion with 2/m point group \cite{Hutchings1972}. We label the [1-10], [11-2], and [111] crystal directions in Fig.~\ref{fig:niogeometry} as $x$, $y$, and $z$, in order to make the indexing of the coefficients simpler.


\subsection{Inverse Faraday and inverse Cotton-Mouton effects}

First, we evaluate the effective magnetic fields $B^\mathrm{eff}$ produced by the opto-magnetic effects according to the expressions shown in Table~\ref{tab:light-matter}. We use the experimental geometry of Ref.~\cite{Tzschaschel2017}, where pulses with full width at half maximum durations of 90~fs at fluences of 80~mJ/cm$^2$, corresponding to a peak electric field of 25~MV/cm, were sent onto a (111)-cut NiO crystal under nearly normal incidence, as sketched in Fig.~\ref{fig:niogeometry}. The pulses with a photon energy of 0.98~eV (well below the $\sim$4.3~eV bandgap \cite{Sawatzky1984}) excited the in-plane magnon mode with a frequency of $\Omega_\mathrm{m}/(2\pi)=0.14~\mathrm{THz}$ through impulsive stimulated Raman scattering.

In this geometry, an excitation of the in-plane magnon mode via the inverse Faraday effect requires the pulse to be circularly polarized in the $xy$ plane, and the corresponding opto-magnetic coefficient is $\alpha_{xyz}$. The $z$-component of the effective magnetic field is given by
\begin{equation}
B^\mathrm{IFE}_z=i\alpha_{xyz}(E_x E_y^\ast - E_y E_x^\ast).
\end{equation}
Here, the circularly polarized electric field of the pulse can be written as $\mathbf{E}=E_0(t)(\mathrm{e}^{i\omega_0t},i\mathrm{e}^{i\omega_0t},0)/\sqrt{2}$, with $E_0(t) = E_0 \mathrm{exp}(-t^2/(2(\tau/\sqrt{8\mathrm{ln}2})^2))$, where $E_0$ is the peak electric field, $\omega_0$ is the center frequency, and $\tau$ is the full width at half maximum duration of the pulse \cite{Juraschek2018}.
For an excitation via the inverse Cotton-Mouton effect, the pulse has to be linearly polarized in the $xy$ plane, and the corresponding coefficient is $\beta'_{xyyx}$. $\beta_{xyyx}$ is zero by symmetry \cite{Tzschaschel2017}. The $z$-component of the effective magnetic field is given by
\begin{equation}
B^\mathrm{ICME}_z=\beta_{xyyx}' (E_x E_y^\ast + E_y E_x^\ast),
\end{equation}
where the linearly polarized electric field is given by $\mathbf{E}=E_0(t)(\mathrm{e}^{i\omega_0t},\mathrm{e}^{i\omega_0t},0)/\sqrt{2}$.


\begin{table}[b]
\centering
\bgroup
\def\arraystretch{1.5}
\caption{
Opto- and phono-magnetic coefficients, and effective magnetic fields calculated in this work. The opto-magnetic coefficients are in units of $\varepsilon_0(V_c/\mu_B)^n$, where $\varepsilon_0$ is the vacuum permittivity, $V_c$ is the unit cell volume, and $n$ is the order of expansion of the dielectric tensor. The phono-magnetic coefficients as displayed here are in units of THz/$\mu_B$ (first order) and (THz$V_c$/$\mu_B$)$^2$ (second order). The effective magnetic fields $B^\mathrm{eff}(\Omega_\mathrm{m})$ are the 0.14 THz components.
}
\begin{tabular}{lll}
\hline\hline
Excitation effect & Coupling~~ & Mag. field \\
\hline
Inverse Faraday & $\alpha_{xyz} = 4\times10^{-3}$ & 28~mT \\
Inverse Cotton-Mouton & $\beta_{xyyx} = 0$ & --  \\
 & $\beta'_{xyyx} = 2.6\times10^{-5}$ & 0.6~mT  \\
\hline
Phonon inverse Faraday~~ & & \\
via phonon orb. mag. mom. & $\bar{a}^\mathrm{POM}_{ijz} = 1.6\times10^{-6}$ & 2.5~mT \\
via crystal electric field~~ & $\bar{a}^\mathrm{CEF}_{ijz} = 6\times10^{-3~\ast}$ & 9.6~T \\
Phonon inverse Cott.-Mout. & $\tilde{b}_{ijzz} = 170$ & --  \\
                   & $\tilde{b}'_{ijxx} = 0$ & -- \\
\multicolumn{3}{l}{$^\ast{}$ Estimate for Tb$_3$Ga$_5$O$_{12}$ from Refs.~\cite{sheng:2006,zhang:2014}} \\
\hline\hline
\end{tabular}
\label{tab:magneticfieldstrengths}
\egroup
\end{table}


\begin{figure*}[t]
\centering
\includegraphics[scale=0.15]{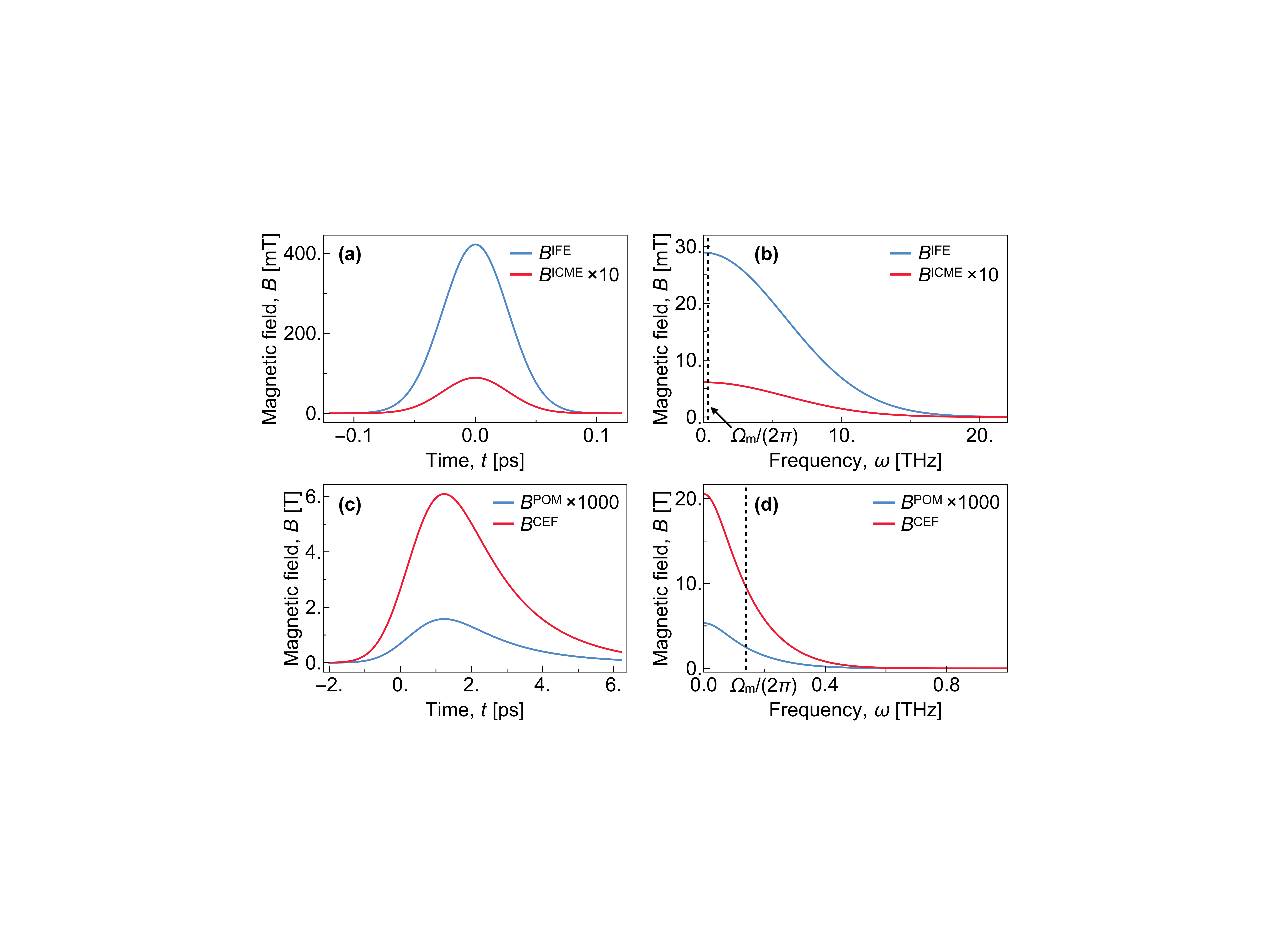}
\caption{
Time- and frequency-dependent effective magnetic fields induced by the opto-magnetic and phono-magnetic effects. (a) $B(t)$ and (b) $B(\omega)$ generated by the inverse Faraday effect (IFE) and inverse Cotton-Mouton effect (ICME) for excitation with a pulse with a photon energy of $\omega_0 \equiv 0.98$~eV, a full width at half maximum duration of $\tau=90$~fs, and a peak electric field of $E_0=25$~MV/cm. (c) $B(t)$ and (d) $B(\omega)$ generated by the phonon inverse Faraday effect and mediated through the phonon orbital magnetic moments (POM) and crystal electric field modulations (CEF) for excitation with a pulse with center frequency $\omega_0=12$~THz, a full width at half maximum duration of $\tau=2.25$~ps, and a peak electric field of $E_0=5$~MV/cm. Note that the top row is in milli tesla, while the bottom row is in tesla. The dashed lines in (b) and (d) mark the frequency of the in-plane magnon mode, $\Omega_\mathrm{m}/(2\pi)=0.14$~THz.
}
\label{fig:magneticfields}
\end{figure*}

We compute the opto-magnetic coefficients at the frequency of the laser pulse $\omega_0$ from first principles as described in the computational details section in the appendix, and we show their values $\alpha_{xyz}(\omega_0)$ and $\beta'_{xyyx}(\omega_0)$ in Table~\ref{tab:magneticfieldstrengths}. The ratio of the two coefficients, $L_0\beta'_{xyyx}(\omega_0)/\alpha_{xyz}(\omega_0)=0.02$, is within the same order of magnitude as the experimental estimate from Ref.~\cite{Tzschaschel2017}. Here, $L_0 = 3.24~\mu_\mathrm{B}/V_\mathrm{c}$ is the magnitude of the equilibrium antiferromagnetic vector obtained from our first-principles calculations. We show the time- and frequency-dependent effective magnetic fields produced by the opto-magnetic effects in Figs.~\ref{fig:magneticfields}(a) and (b). The driving force of the magnon mode is the Fourier component of the effective magnetic field that is resonant with its eigenfrequency. We find the 0.14~THz components of the effective magnetic fields to yield $B^\mathrm{IFE}(\Omega_\mathrm{m})=28$~mT and $B^\mathrm{ICME}(\Omega_\mathrm{m})=0.6$~mT, listed in Table~\ref{tab:magneticfieldstrengths}. Note that the effective magnetic field determines the induced magnetization, which is larger for the inverse Faraday effect than for the inverse Cotton-Mouton effect. This can however be overcompensated by the anisotropy of the system, and the anisotropy factor in NiO is large enough so that the inverse Cotton-Mouton effect ends up dominating over the inverse Faraday effect.


\subsection{Phonon inverse Faraday and phonon inverse Cotton-Mouton effects}

Second, we evaluate the effective magnetic fields produced by the phono-magnetic effects according to the expressions shown in Table~\ref{tab:phonon-matter}, using the same geometry as before. For the excitation of the in-plane magnon mode via the phonon inverse Faraday effect, the infrared-active $B_\mathrm{u}$ modes perpendicular to the [111] direction have to be coherently excited with a circularly polarized mid-infrared pulse. The corresponding phono-magnetic coefficient is $\bar{a}_{ijz}$, where $i$ and $j$ now denote the nearly degenerate $B_\mathrm{u}$ modes at 12.06 and 12.08~THz. The $z$-component of the effective magnetic field is given by
\begin{equation}
B^\mathrm{PIFE}_z=\bar{a}_{xyz}(Q_i \dot{Q}_j - Q_j \dot{Q}_i).
\end{equation}
For an excitation via the phonon inverse Cotton-Mouton effect, one of the $B_\mathrm{u}$ modes has to be excited with a linearly polarized mid-infrared pulse. The corresponding phono-magnetic coefficients are $b_{iizz}$ and $b'_{iixx}$ (equivalently for $j$). As we will see, the phonon inverse Cotton-Mouton effect is zero in NiO.

For the phonon inverse Faraday effect, no straightforward first-principles implementation to calculate the imaginary part of the dynamical matrix exists to date, in contrast to the imaginary part of the dielectric tensor that is necessary for the inverse Faraday effect. We identified two different microscopic origins for the phono-magnetic coefficient $\bar{a}_{ijz}$, and we will evaluate them separately, $\bar{a}_{ijz}=\bar{a}_{ijz}^\mathrm{POM}+\bar{a}_{ijz}^\mathrm{CEF}$. The first contribution $\bar{a}_{ijz}^\mathrm{POM}$ comes from the coupling of the orbital magnetic moment produced by the coherently excited circularly polarized phonon modes to the spin, which has previously been described in the context of the dynamical multiferroic effect, and we evaluate it according to the formalism developed in Refs.~\cite{juraschek2:2017,Juraschek2019}, as outlined in the computational details section. In the second contribution, $\bar{a}_{ijz}^\mathrm{CEF}$, the coherently excited phonons couple to the spin through a modulation of the crystal electric field around the magnetic ion. This modulation of the crystal field is also at the origin of the Raman process of spin relaxation \cite{VanVleck1940,Orbach1961,Ray1967,Ioselevich1995} and the phonon Hall effect \cite{sheng:2006,Zhang2011}, in which incoherent (thermal) phonons interact with the spin. $\bar{a}_{ijz}^\mathrm{CEF}$ can in principle be estimated using crystal electric field theory with input parameters calculated with quantum chemistry methods, which is beyond the scope of this study. Experimental data on this type of spin-phonon coupling is scarce, and we use the (to our knowledge) only literature value for a transition metal oxide, Tb$_3$Ga$_5$O$_{12}$, from Refs.~\cite{sheng:2006,zhang:2014}.

The coefficients of the phonon inverse Cotton-Mouton effect were computed analogously to the opto-magnetic effects as described in the computational details section. The spin-phonon coupling here includes all microscopic contributions from modulations of exchange interactions and anisotropies that are relevant to the ionic Raman scattering process. We list the phono-magnetic coefficients $\bar{a}_{ijz}^\mathrm{POM}$, $\bar{a}_{ijz}^\mathrm{CEF}$, $\tilde{b}_{iizz}$, and $\tilde{b}'_{iixx}$ in Table~\ref{tab:magneticfieldstrengths}.

To obtain the coherent phonon amplitudes $Q_i$ and $Q_j$, we numerically solve the equation of motion that models the coherent excitation of the $B_\mathrm{u}$ modes by a mid-infrared pulse:
\begin{eqnarray}\label{eq:oscillatormodel}
\ddot{Q}_i + \kappa_i \dot{Q}_i + \Omega_i^2 Q_i = Z_i E_i(t).
\end{eqnarray}
Here, $\Omega_i$ are the eigenfrequencies, $\kappa_i$ the linewidths, and $Z_i$ the mode effective charges of the $B_\mathrm{u}$ modes, and $E_i(t)$ is the electric-field component of the pulse \cite{Juraschek2018}. (The equation for the $j$ component is equivalent.) In order to excite circularly polarized phonon modes, the mid-infrared pulse has to be circularly polarized and of the form $(E_i(t),E_j(t))=(E(t),E(t-2\pi/(4\omega_0)))$, where $E(t) = E_0 \mathrm{exp}(-t^2/(2(\tau/\sqrt{8\mathrm{ln}2})^2))\cos(\omega_0 t)$. We simulate a mid-infrared pulse that is resonant with the phonon modes, $\omega_0 = \Omega_i \approx \Omega_j$, with $\tau=2.25$~ps and $E_0=5$~MV/cm. This pulse has the same energy as that used to model the opto-magnetic effects. According to the Lindemann stability criterion, a material will melt when the root-mean square displacements of the atoms exceed 10-20\%{} of the interatomic distance. For the pulse used here, we expect a root-mean square displacement of the oxygen ions of around 0.4\ang{}, which corresponds to 20\%{} of the interatomic distance. Therefore, the values presented here should be seen as an upper limit for the induced magnetic fields. We show the time- and frequency-dependent effective magnetic fields produced by the phono-magnetic effects in Figs.~\ref{fig:magneticfields}(c) and (d). We find the effective magnetic field components at the frequency of the magnon mode to yield $B^\mathrm{POM}(\Omega_\mathrm{m})=2.5$~mT and $B^\mathrm{CEF}(\Omega_\mathrm{m})=9.6$~T, listed in Table~\ref{tab:magneticfieldstrengths}. The phonon inverse Faraday effect mediated through the orbital magnetic moments of phonons has the same order of magnitude as the opto-magnetic effects for comparable pulse energies. Using the estimated value for the crystal electric field modulation, we calculate that the phonon inverse Faraday effect would yield an effective magnetic field that is three orders of magnitude larger than the other effects. The coefficient of the phonon inverse Cotton-Mouton effect for the expansion in $L^2$ is zero. In contrast, the coefficient for the expansion in $M^2$ is nonzero, but as the effective magnetic field scales with the equilibrium ferromagnetic vector, $M_0=0$, this effect also vanishes. We therefore expect no second-order phono-magnetic effect for the in-plane magnon mode in NiO.


\subsection{Discussion}

We have shown that for comparable pulse energies, the phono-magnetic effects that are based on ionic Raman scattering generate effective magnetic fields that are comparable to those generated by the opto-magnetic effects that are based on impulsive stimulated Raman scattering. The pulse frequencies required to coherently excite phonons for the phono-magnetic effects lie in the terahertz and mid-infrared region ($\sim$tens of meV), where parasitic electronic excitations are reduced compared to the pulses commonly used to trigger opto-magnetic effects, which usually use visible light ($\sim$eV). This makes the phono-magnetic effects more selective than the opto-magnetic effects and opens up the possibility of generating coherent magnons in the electronic ground state in small band gap semiconductors. While in NiO, there is only one pair of infrared-active phonon modes present that is relevant for the coupling to the magnon, materials like yttrium iron garnet host a large variety of different phonon and magnon modes. Here, highly selective phonon excitation, such as that recently demonstrated in a high-temperature superconductor \cite{liu2020}, may play a prominent role in the phono-magnetic mechanisms.


\begin{figure}[t]
\centering
\includegraphics[scale=0.145]{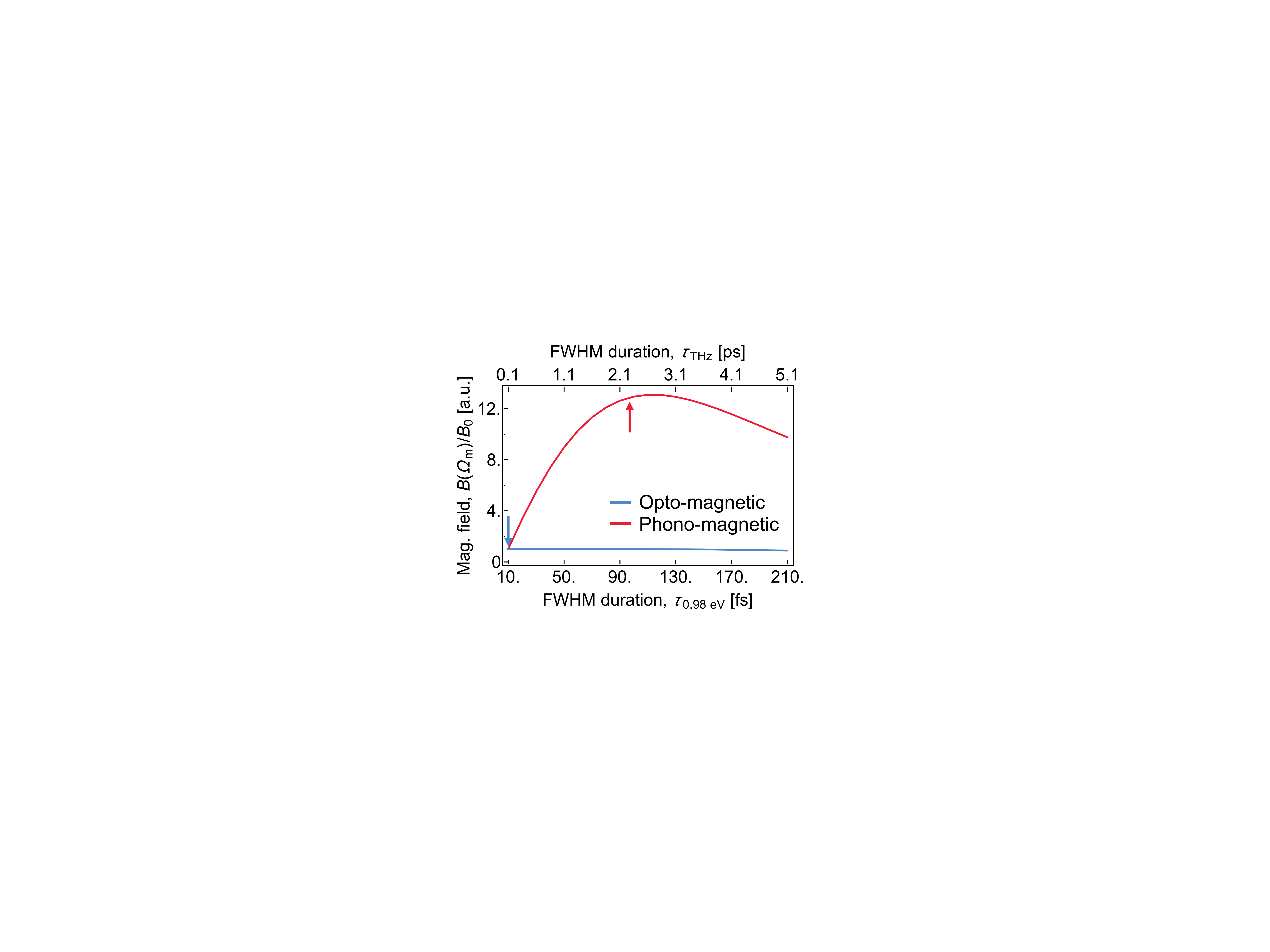}
\caption{
Magnetic field components at the frequency of the in-plane magnon mode, $B(\Omega_\mathrm{m})$, for varying durations and constant energies (peak electric field adjusted accordingly) of the 0.98~eV and terahertz pulses. The values are normalized to the respective values of the shortest pulses, $B_0$. Arrows mark the values of the pulse duration used for the calculations of the effective magnetic fields in the main text.
}
\label{fig:pulseduration}
\end{figure}

To estimate the strength of the phonon inverse Faraday effect mediated through the crystal electric field, we turned to literature values for the coupling. To our knowledge, only a few values have been reported, specifically for 4$f$ paramagnetic Tb$_3$Ga$_5$O$_{12}$ \cite{sheng:2006,zhang:2014} and some rare-earth trihalides \cite{schaack:1976,schaack:1977,zhang:2014,Juraschek2020}. We acknowledge, however, that the real coupling strength may strongly differ for materials with 3$d$ and 4$f$ magnetism and realistic values for NiO should be computed from first-principles in future work. Nevertheless, this coupling mechanism yields the largest effective magnetic field for all mechanisms. As an example, the spin-precessional amplitudes induced by a circular phononic drive in the weak ferromagnet ErFeO$_3$ \cite{nova:2017} are larger by an order of magnitude than that by a circular photonic drive \cite{DeJong2011} for comparable total pulse energies. This comparison and our large calculated value for the example coupling strength suggest that crystal electric field modulation could be a significant and potentially the dominant microscopic mechanism responsible for the magnon excitation in Ref.~\cite{nova:2017}, in addition to the originally proposed orbital magnetic moments of phonons \cite{nova:2017,juraschek2:2017}. Quantitative calculations of the coupling will be necessary in the future in order to confirm this statement and to predict materials in which the effect can be maximized. In addition, it should be noted that the opto-magnetic effects may benefit from resonant enhancement when the photon energy of the pulse is tuned to match electronic transitions in the vicinity of the band gap \cite{Iida2011}, which will have to be taken into account in a rigorous quantitative comparison of opto- and phono-magnetic effects. We see no second-order phono-magnetic effect for antiferromagnetic NiO, in contrast to the well-known second-order opto-magnetic inverse Cotton-Mouton effect, and despite the observation that the frequencies of some phonon modes depend on the magnetic order \cite{Kant2009,Aytan2017}.

Finally, we discuss the effect of the pulse duration on the effective magnetic field. We show $B(\Omega_\mathrm{m})$ for different values of the full width at half maximum pulse duration $\tau$ in Fig.~\ref{fig:pulseduration}, where we adjust the peak electric field $E_0$ such that the pulse energy remains constant. The data sets for the 0.98~eV and mid-infrared pulses are normalized to the value of the shortest pulse. In the opto-magnetic effects, the 0.14~THz component of the effective magnetic field is nearly independent of $\tau$ for the 10~fs to 200~fs range shown here, because $\Omega_\mathrm{m}\ll\omega_0$ can be regarded as an effectively static component of the difference-frequency mixing of the electric-field components in the pulse. In the phono-magnetic effects in contrast, the effective magnetic field depends strongly on the pulse duration, because the amplitude of the $B_\mathrm{u}$ modes increases for longer resonant driving pulses. This dependence on the pulse duration makes it possible to fine-tune the efficiency of the phono-magnetic effects. For very short pulses, additional driving mechanisms for the spins come into play that cannot be captured with the effective magnetic field formalism presented here and require, in principle, a fully quantum mechanical approach \cite{popova:2011,Berritta2016,Majedi2020}. In addition to the paramagnetic response of the spins to the effective magnetic field, an unquenching of electronic orbital magnetic moments produces a diamagnetic response through optically induced virtual transitions that may even dominate the magnetic response on femtosecond timescales \cite{reid:2010,Mikhaylovskiy2012}. For the opto-magnetic inverse Faraday effect, the effective magnetic field formalism has been shown to hold on the timescale of nanoseconds \cite{vanderziel:1965}. The picosecond timescale of the phono-magnetic effects lies right in between these two limits. We therefore anticipate that the effective magnetic field formalism is a reasonable approach, however additional effects should be expected in an experiment. These will have to be taken into account in order to make quantitatively accurate predictions in the future.

As tabletop terahertz and mid-infrared sources are becoming widely available \cite{Liu2017}, we anticipate that an increasing number of phono-magnetic phenomena will be explored over the course of the next few years. The theory presented here lays the foundation for this emerging phononic approach to antiferromagnetic spintronics.


\begin{acknowledgments}
We are grateful to T. Kampfrath (Fritz Haber Institute), S. Maehrlein (Columbia University), J. Lehmann, S. Pal, and M. Fiebig (ETH Zurich), R. Pisarev and A. Kalashnikova (Ioffe Institute), R. Merlin (University of Michigan), M. Fechner and A. Rubio (MPSD Hamburg), and C. Tzschaschel (Harvard University) for useful discussions. This work was supported by the Swiss National Science Foundation (SNSF) under Project ID 184259, the DARPA DRINQS Program under Award No. D18AC00014 at Harvard University, and the European Research Council (ERC) under the European Union’s Horizon 2020 research and innovation programme under Grant Agreement No. 810451. Calculations were performed at the Swiss National Supercomputing Centre (CSCS) supported by the Project IDs s624, p504, and s889.
\end{acknowledgments}


\begin{figure}[b]
\centering
\includegraphics[scale=0.095]{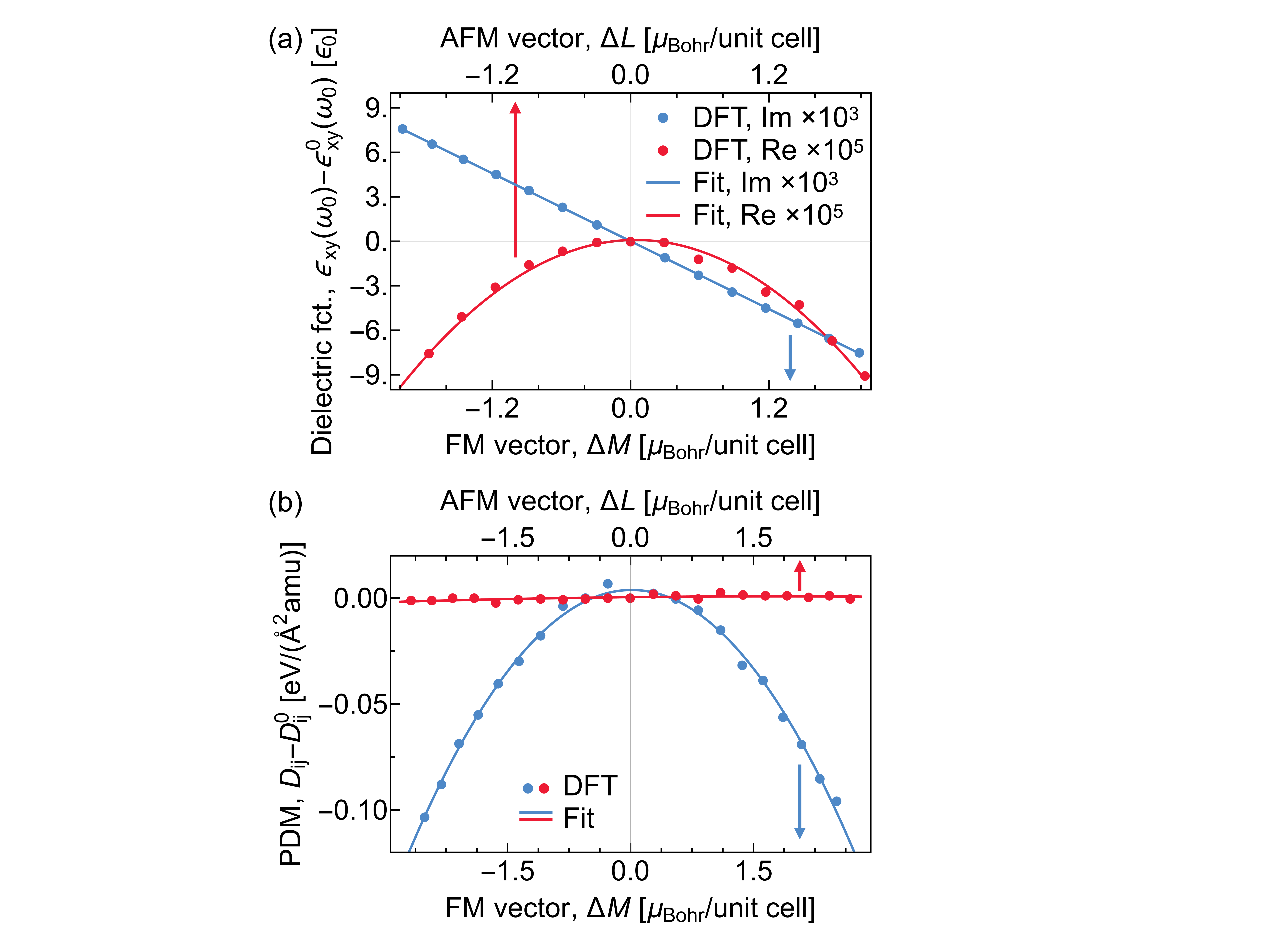}
\caption{
Opto- and phono-magnetic coefficients from density functional theory. (a) Relevant real and imaginary parts of the dielectric tensor as function of the ferro and antiferromagnetic vector components. The dielectric tensor is given in units of the vacuum permittivity $\varepsilon_0$ and evaluated at the frequency of the laser pulse $\hbar\omega_0=0.98$~eV. We subtract the equilibrium value. (b) Projected dynamical matrix for the for the in-plane $B_\mathrm{u}$ phonon modes $i$ and $j$ perpendicular to the [111] direction as function of the ferro and antiferromagnetic vectors. The dynamical matrix is given in units of eV/(\ang$^2$amu). We subtract the equilibrium value.
}
\label{fig:dftcoefficients}
\end{figure}


\begin{appendix}

\section*{Appendix: Computational details}

We calculate the opto- and phono-magnetic coefficients, phonon eigenvectors, and phonon eigenfrequencies from first-principles using the density functional theory formalism as implemented in the Vienna Ab initio Simulation Package (VASP) \cite{kresse:1996,kresse2:1996} and the frozen-phonon method as implemented in the Phonopy package \cite{phonopy}. We use the VASP projector-augmented wave (PAW) pseudopotentials with valencies Ni 3p$^6$3d$^9$4s$^1$ and O 2s$^2$2p$^4$ and converge the Hellmann-Feynman forces to 25~$\mu$eV/\ang{} using a plane-wave energy cut-off of 800~eV and a 8$\times$8$\times$8 gamma-centered $k$-point mesh to sample the Brillouin zone. We take spin-orbit coupling into account for every calculation. For the exchange-correlation functional, we choose the Perdew-Burke-Ernzerhof revised for solids (PBEsol) form of the generalized gradient approximation (GGA) \cite{csonka:2009}. We apply an on-site Coulomb interaction of 4~eV on the Ni 3$d$ states that well reproduces both the $G$-type antiferromagnetic ordering and lattice dynamical properties \cite{Reichardt1975}. The volume of the fully relaxed primitive unit cell is $V_\mathrm{c} = 35.59~\mathrm{\AA}^3$ and the magnitude of the equilibrium antiferromagnetic vector projected into Wigner-Seitz spheres $L_0 = 3.24~\mu_\mathrm{B}/V_\mathrm{c}$. We find the mode effective charges of the infrared-active $B_\mathrm{u}$ modes to be 0.84$e$, where $e$ is the elementary charge \cite{Gonze1997}, and we use common values for the phonon linewidths in oxides, $\kappa_i\approx0.05\times\Omega_i/(2\pi)$ \cite{juraschek:2017}.

We vary the angle of the noncollinear spins of NiO along the directions of the in-plane magnon mode coordinates as displayed in Fig.~\ref{fig:niogeometry} in increments of 5$^\circ$ between $\pm$90$^\circ$ and calculate the frequency-dependent dielectric function $\varepsilon_{ij}$ and the projected dynamical matrix $D_{ij}$ for each step. We then fit the functions $\varepsilon_{ij}(\mathbf{M},\mathbf{L})$ and $D_{ij}(\mathbf{M},\mathbf{L})$ to quadratic polynomials in $\mathbf{M}$ and $\mathbf{L}$ as shown in Fig.~\ref{fig:dftcoefficients} to obtain the magneto-optical and magneto-phononic coefficients. A similar method has been used to obtain nonlinear phonon-phonon couplings in a recent study \cite{Khalsa2018}. Note that the coefficients for the phonon (inverse) Cotton-Mouton effects can, in principle, also be obtained by calculating the phonon-amplitude dependent exchange interactions and anisotropies instead, see, e.g., Ref.~\cite{Fechner2018}. The phono-magnetic coefficient, $\bar{a}_{ijz}^\mathrm{POM}$, can be regarded as a gyromagnetic ratio of the circularly polarized phonon modes $i$ and $j$, which is given by $a_{ijz} = \sum_{n} \gamma_{n} \mathbf{q}_{ni} \times \mathbf{q}_{nj}$, where $\mathbf{q}_{ni}$ are the eigenvectors of phonon mode $i$ corresponding to ion $n$, and the sum runs over all ions in the unit cell. $\gamma_{n}=e\mathbf{Z}^{\ast}_{n}/(2 \mathcal{M}_{n})$ is the gyromagnetic ratio of ion $n$, where $e$ is the elementary charge, $\mathbf{Z}^{\ast}_{n}$ is the Born effective charge tensor and $\mathcal{M}_{n}$ is the atomic mass. For $\bar{a}_{ijz}^\mathrm{CEF}$, we use the literature value for Tb$_3$Ga$_5$O$_{12}$ provided in Refs.~\cite{sheng:2006,zhang:2014}.

\end{appendix}



%




\end{document}